\documentclass[a4paper,english,cits]{PoS}
\usepackage[T1]{fontenc}
\usepackage[latin1]{inputenc}
\usepackage{rotating}
\usepackage{graphicx}
\usepackage{amssymb}
\usepackage[authoryear]{natbib}

\makeatletter

\providecommand{\tabularnewline}{\\}

\def\jnl@style{\it}
\def\aaref@jnl#1{{\jnl@style#1}}

\def\aaref@jnl#1{{\jnl@style#1}}

\def\aj{\aaref@jnl{AJ}}                   
\def\araa{\aaref@jnl{ARA\&A}}             
\def\apj{\aaref@jnl{ApJ}}                 
\def\apjl{\aaref@jnl{ApJ}}                
\def\apjs{\aaref@jnl{ApJS}}               
\def\ao{\aaref@jnl{Appl.~Opt.}}           
\def\apss{\aaref@jnl{Ap\&SS}}             
\def\aap{\aaref@jnl{A\&A}}                
\def\aapr{\aaref@jnl{A\&A~Rev.}}          
\def\aaps{\aaref@jnl{A\&AS}}              
\def\azh{\aaref@jnl{AZh}}                 
\def\baas{\aaref@jnl{BAAS}}               
\def\jrasc{\aaref@jnl{JRASC}}             
\def\memras{\aaref@jnl{MmRAS}}            
\def\mnras{\aaref@jnl{MNRAS}}             
\def\pra{\aaref@jnl{Phys.~Rev.~A}}        
\def\prb{\aaref@jnl{Phys.~Rev.~B}}        
\def\prc{\aaref@jnl{Phys.~Rev.~C}}        
\def\prd{\aaref@jnl{Phys.~Rev.~D}}        
\def\pre{\aaref@jnl{Phys.~Rev.~E}}        
\def\prl{\aaref@jnl{Phys.~Rev.~Lett.}}    
\def\pasp{\aaref@jnl{PASP}}               
\def\pasj{\aaref@jnl{PASJ}}               
\def\qjras{\aaref@jnl{QJRAS}}             
\def\skytel{\aaref@jnl{S\&T}}             
\def\solphys{\aaref@jnl{Sol.~Phys.}}      
\def\sovast{\aaref@jnl{Soviet~Ast.}}      
\def\ssr{\aaref@jnl{Space~Sci.~Rev.}}     
\def\zap{\aaref@jnl{ZAp}}                 
\def\nat{\aaref@jnl{Nature}}              
\def\iaucirc{\aaref@jnl{IAU~Circ.}}       
\def\aplett{\aaref@jnl{Astrophys.~Lett.}} 
\def\apspr{\aaref@jnl{Astrophys.~Space~Phys.~Res.}}
\def\bain{\aaref@jnl{Bull.~Astron.~Inst.~Netherlands}} 
\def\fcp{\aaref@jnl{Fund.~Cosmic~Phys.}}  
\def\gca{\aaref@jnl{Geochim.~Cosmochim.~Acta}}   
\def\grl{\aaref@jnl{Geophys.~Res.~Lett.}} 
\def\jcp{\aaref@jnl{J.~Chem.~Phys.}}      
\def\jgr{\aaref@jnl{J.~Geophys.~Res.}}    
\def\jqsrt{\aaref@jnl{J.~Quant.~Spec.~Radiat.~Transf.}}
\def\memsai{\aaref@jnl{Mem.~Soc.~Astron.~Italiana}}
\def\nphysa{\aaref@jnl{Nucl.~Phys.~A}}   
\def\physrep{\aaref@jnl{Phys.~Rep.}}   
\def\physscr{\aaref@jnl{Phys.~Scr}}   
\def\planss{\aaref@jnl{Planet.~Space~Sci.}}   
\def\procspie{\aaref@jnl{Proc.~SPIE}}   

\usepackage{babel}
\makeatother

\title{On burning regimes and long duration X-ray bursts}
\ShortTitle{On burning regimes and long duration X-ray bursts}

\author{\speaker{L. Keek}\\
SRON Netherlands Institute for Space Research, Sorbonnelaan 2,
NL - 3584 CA Utrecht, the Netherlands\\
Astronomical Institute, Utrecht University, P.O. Box 80000, NL
- 3508 TA Utrecht, the Netherlands\\
E-mail: \email{l.keek@sron.nl}}

\author{J.\,J.\,M. in 't Zand\\
SRON Netherlands Institute for Space Research, Sorbonnelaan 2,
NL - 3584 CA Utrecht, the Netherlands}

\abstract{
Hydrogen and helium accreted onto a neutron star undergo thermonuclear
burning. Explosive burning is observed as a type I X-ray burst. We
describe the different burning regimes and focus on some of the current
inconsistencies between theory and observations. Of special interest
are the rare kinds of X-ray bursts such as carbon-fueled superbursts
and helium-fueled intermediately long X-ray bursts. These bursts are
thought to originate deeper in the neutron star envelope, such that
they are probes of the thermal properties of the crust. We investigate
the possibility of observing superbursts with the wide-field instruments
INTEGRAL-ISGRI and Swift-BAT. We find that only the brightest bursts
are detectable.}

\FullConference{7th INTEGRAL Workshop\\
		 September 8-11 2008\\
		 Copenhagen, Denmark}

\begin{document}

\section{Introduction}

Neutron stars in low-mass X-ray binaries can accrete material containing
hydrogen and helium from a lower-mass companion star. Due to the high
surface gravity of neutron stars, the material is compressed within
a matter of hours to such high densities that, when it is just a few
meters thick, the conditions for fusion of hydrogen and of helium
are reached. If the thermonuclear fusion is unstable, the entire surface
is burned within one second. Afterwards, the neutron star envelope
cools down on a typical timescale of $10\,\mathrm{s}$. This is observable
as a type I X-ray burst (\citealt{Woosley1976,Maraschi1977,Lamb1978}).
The spectrum is usually well-fit by a simple black body with a peak
temperature of $2$ to $3\,\mathrm{keV}$, which makes them detectable
in the classical X-ray band. During the tail of the burst, the black
body temperature decreases. Since their discovery (\citealt{Grindlay1976}),
several thousands of these bursts have been observed from approximately
90 sources (e.g., \citealt{2004ZandWFC,Galloway2007,intZand2007}; for reviews
see \citealt{Lewin1993,Strohmayer2006}). In this proceedings paper
we focus on aspects of the burning processes responsible for X-ray
bursts and on rare kinds of long bursts that have been discovered
in recent years.

\section{Stable and unstable burning regimes}

Several nuclear reaction chains are responsible for the burning on a
hydrogen/helium accreting neutron star. For instance, hydrogen
burns to helium via the CNO cycle and helium burns to carbon through
the triple-alpha process. The reaction rates of these processes can
be strongly dependent on the temperature in the envelope, such that
a small increase in the temperature causes a large rise in the energy
generation rate. If the cooling rate is too low, this again increases the temperature, leading
to a thermonuclear runaway, which is observable as an X-ray burst.
At higher temperatures the dependence weakens, such that above a certain
temperature burning proceeds in a stable manner. For hydrogen burning
the transition is at $T\simeq8\cdot10^{7}\,\mathrm{K}$, while for
helium burning it is $T\simeq3.5\cdot10^{8}\,\mathrm{K}$ (\citealt{Bildsten1998}).
Above this treshold, hydrogen burning follows a slightly different
reaction chain known as the hot CNO cycle, which is temperature independent.

The temperature in the envelope is influenced by heating as well as
cooling processes and increases with depth. The most important source
of heat is stable thermonuclear burning of hydrogen and helium in
the envelope itself. Cooling is mostly radiative. Part of the heat
may be conducted inward, but in the envelope this is estimated to
be only a small fraction (\citealt{Eichler1989}). Another source
of heat is the crust, a solid layer located below the envelope. During
accretion the crust is compressed, which induces pycnonuclear reactions
and electron captures. About 10\% of the thus generated energy heats
the envelope, while the rest is conducted deeper into the neutron
star. The heat flux generated per accreted nucleon is thought to be
$Q=0.15\,\mathrm{MeV\, nucleon^{-1}}$ (\citealt{Haensel2003}), although
a recent study indicates it may be a few times higher (\citealt{Gupta2007}).

The amount of heating depends on the mass accretion rate $\dot{M}$.
For a higher $\dot{M}$ the heating by processes in the crust is stronger, leading to a higher temperature at the base
of the burning layer. This influences the stability of thermonuclear
burning, such that several burning regimes can be distinguished as
a function of $\dot{M}$. %
\begin{table}

\caption{\label{tab:Burst-regimes-as}Burning regimes as a function of mass
accretion rate (\citealt{Fujimoto1981,Bildsten1998}), for a hydrogen/helium accreting neutron
star mass of $1.4\mathrm{M_{\odot}}$, a radius of$10\,\mathrm{km}$,
a hydrogen mass fraction $X=0.7$ and a CNO mass fraction $Z_{\mathrm{CNO}}=0.01$.}

\begin{centering}\begin{tabular}{lll}
\hline 
Regime&
$\dot{M}/\dot{M}_{\mathrm{Edd}}$&
Burning\tabularnewline
\hline
I&
&
Mixed H/He flash (H ignites first)\tabularnewline
&
$0.5\%$&
\tabularnewline
II&
&
He flash (stable H burning)\tabularnewline
&
$3\%$&
\tabularnewline
III&
&
Mixed H/He flash (He ignites first)\tabularnewline
&
$100\%$&
\tabularnewline
IV&
&
Stable H/He burning\tabularnewline
\hline
\end{tabular}\par\end{centering}
\end{table}
 In Table \ref{tab:Burst-regimes-as} we indicate the mass accretion
rate at the transition between the regimes as found in models (\citealt{Fujimoto1981,Bildsten1998}).
All regimes are observed, but the mass accretion rate at the transitions
is only approximately in agreement with the theoretical picture. For
example, \citet{Cornelisse2003} observe the transition from regime
I to II at $\dot{M}\simeq0.1\dot{M}_{\mathrm{Edd}}$ (note that
\citealt{1826:galloway04apj} argue this is the transition from
regime II to III). \citet{Paradijs1988}
find indications of an increasing amount of stable H/He burning at
$\dot{M}\gtrsim0.1\dot{M}_{\mathrm{Edd}}$, suggesting that there
is a smooth transition from regime III to IV starting at a lower accretion
rate than predicted by theory. The theoretical picture changes somewhat
if certain important effects are incorporated in the models. In their
models that include the effect of sedimentation \citet{Peng2007}
find at low accretion rates, $0.3\%\lesssim\dot{M}/\dot{M}_{\mathrm{Edd}}\lesssim1\%$,
a new regime of pure hydrogen bursts, where helium sinks to larger
depths (see also \citealt{Cooper2007a,intZand2007}). Another important
effect is turbulent mixing due to rotation (\citealt{Fujimoto1993}).
When matter is accreted from the disk onto the neutron star it likely
has a higher angular velocity than the neutron star itself. The accreted
angular momentum is shared with the rest of the star. This leads to
differential rotation, which causes turbulence that can mix the chemical
composition of neighboring layers. If mixing is important, hydrogen
and helium are diffused to larger depths, where the temperature is
higher. This shifts the boundaries of the predicted burning regimes
(Table \ref{tab:Burst-regimes-as}). \citet{Yoon2004} investigated
the effect of rotationally induced mixing on the stability of burning
in the case of helium accreting white dwarfs. They find that the effect
is important, as it leads to increased stability of thermonuclear
burning. \citet{Piro2007} study helium accretion on neutron stars
and find that also in this case mixing increases stability. Their
models include a rotationally induced magnetic field, which very efficiently
transports angular momentum through the envelope. Recently \citealt{Keek2008a}
studied stable burning of helium in the neutron star envelope using
a one dimensional hydrodynamic stellar evolution code including the
effect of both rotational mixing and a rotationally induced magnetic
field. They find that stability may be increased even further than
\citet{Piro2007} predict, as the stabilizing effect is substantial
for models at all rotation rates. Combined with the increased heat
flux from the crust predicted by \citet{Gupta2007}, these effects
may enable us to better explain the observed burning behavior as a
function of mass accretion rate.

\section{Burst duration}

From our overview of the burning regimes (Table \ref{tab:Burst-regimes-as}),
we see that flashes may occur in burning layers that have different relative hydrogen and helium
contents. Observationally we can discriminate between these bursts
using their duration. The duration of a burst is set by the cooling
time scale of the burning layer, which is determined by the depth
at which the burning takes place. Typically this is of the order of
$10\,\mathrm{s}$. In the presence of hydrogen, however, prolonged
nuclear burning can extend the burst duration up to around $100\,\mathrm{s}$.
A series of proton captures and $\beta$-decays creates heavy elements
with mass numbers of up to 100 (e.g., \citealt{2003Schatz,Fisker2008}).
This is known as the rp-process. Therefore, bursts last longer when
the fuel has a larger hydrogen content. In a histogram of the burst
decay time for a large number of type I bursts (see \citealt{Zand2007seattle}
or \citealt{Chenevez2008} in these proceedings) we see a broad distribution
of bursts with exponential decay times of a few seconds up to $100\,\mathrm{s}$.
Surprisingly, there are also two groups of bursts with much longer
decay times of up to $10^{4}\,\mathrm{s}$: the so-called intermediate
bursts and the superbursts. Their decay times are too long to be explained
by prolonged burning through the rp-process. Therefore, these represent
the cooling time scales of very thick layers. The thicker layers imply
higher ignition pressures and, thus, different ignition temperatures
or fuel compositions. Indeed, intermediately long bursts are thought
to be due to much thicker hydrogen/helium layers on colder neutron
stars (\citealt{Zand2005,Cumming2006}) and superbursts are thought
to result from carbon burning (\citealt{Cornelisse2000,Cumming2001,Strohmayer2002}).
\begin{table}

\caption{\label{tab:Comparisson-of-the}Comparison of the typical properties
of different kinds of type I bursts.}

\begin{centering}\begin{tabular}{llll}
\hline 
&
normal&
intermediate&
superburst\tabularnewline
\hline
duration&
$10-100\,$seconds&
$15-40\,$minutes&
$1\,$day\tabularnewline
fluence&
$10^{39}\,\mathrm{erg}$&
$10^{40}-10^{41}\,\mathrm{erg}$&
$10^{42}\,\mathrm{erg}$\tabularnewline
recurrence time&
hours--days&
tens of days&
$1\,$year\tabularnewline
number observed&
$1000$s &
$\sim20$ &
$15$ \tabularnewline
&
from $\sim90$ sources&
from $8$ sources&
from $10$ sources\tabularnewline
\hline
\end{tabular}\par\end{centering}
\end{table}
 Comparing the properties of these different kinds of bursts in Table
\ref{tab:Comparisson-of-the}, we see that they are indeed consistent
with the burning of thicker layers of fuel in the longer bursts, as
their fluence is also higher. Since it takes a longer time to accumulate
a thicker layer, the intermediate and superbursts have longer recurrence
times, leading to a much lower number of observed instances than for
normal bursts. For example, only $15$ superbursts have been observed
from $10$ sources (Table \ref{tab:Overview-of-all}). Therefore,
most long duration bursts were observed relatively recently, after
the launch of the observatories BeppoSAX and RXTE, which collected
unprecedented amounts of exposure time on many of the bursting sources
in our Galaxy.

\section{Intermediate bursts}

Most intermediate bursts are observed from so-called ultra-compact
X-ray binaries (UCXBs), which have an orbital period of less than
$80$ minutes. In such a small orbit only a companion star fits which
has lost its hydrogen-rich envelope. The neutron star, therefore,
likely does not accrete any hydrogen, but may accrete helium (\citealt{Zand2005}).
Without stable hydrogen burning, the envelope is cold, such that a
much thicker layer has to be accreted before the ignition conditions
for helium burning are reached (\citealt{Cumming2006}). Furthermore,
the most important source of heat is in this case crustal heating.
This makes intermediate bursts probes of the neutron star crust while
mass accretion is ongoing.

A thick helium layer could also form on a hydrogen-accreting neutron
star, if at low mass accretion rates pure hydrogen bursts occur, while
sedimentation causes helium to settle at larger depths (\citealt{Peng2007}).
The hydrogen bursts would be too weak to ignite the helium layer.
Furthermore, these bursts may not exceed the accretion luminosity
and may therefore not be detectable. Two intermediate bursts have been
argued to be due to this scenario: a burst from the globular cluster M28 (\citealt{Gotthelf1997,Peng2007}) and one from IGR~17254-3257 (\citealt{Chenevez2007}). \citealt{intZand2007}
point out some issues which may prevent it from occurring.

Intermediately long bursts with durations of 6 to 25 minutes have
also been observed from GX~17+2 (\citealt{2002Kuulkers}). These are  interesting cases, since the source is not known to be an UCXB and it accretes at near-Eddington rates, while most UCXBs accrete at relatively low rates.

\section{Questions about superbursts}

\begin{figure}
\begin{centering}\includegraphics[bb=30bp 200bp 580bp 560bp,clip,width=0.8\columnwidth]{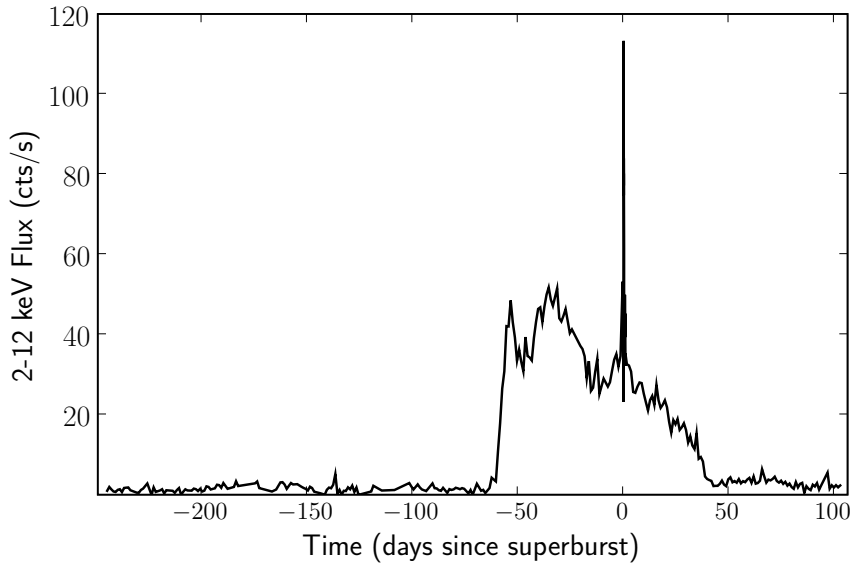}\par\end{centering}

\caption{\label{fig:Superburst-from-4U}RXTE ASM light curve of 4U~1608-522 during the 2005 major outburst. 55 days after the start of the outburst a superburst occurred (\citealt{Keek2008}). Time resolution is two weeks, except during
the superburst where the 90\,s ASM dwells are used.}
\end{figure}
\begin{sidewaystable}

\caption{\label{tab:Overview-of-all}Overview of the ten known superbursters
in order of increasing exponential decay time $\tau_{\mathrm{exp}}$
(adapted from \citealt{Kuulkers2003a}).}

\renewcommand{\tabcolsep}{0.2pc} 
\renewcommand{\arraystretch}{1.2} 
\begin{tabular}{lcccccccccc}
\hline
source                                               & {\bf GX 17+2} & {\bf 1820$-$303} & {\bf 1636$-$536} & {\bf Ser~X-1} & {\bf 1735$-$444} & {\bf GX\,3+1} & {\bf 0614+091}$^b$ & {\bf 1731$-$260} & {\bf 1608-522} & {\bf 1254-690}\\
\hline
number of superbursts & 4 & 1 & 3 & 1 & 1 & 1 & 1 & 1 & 1 & 1\\
onset observed                                       & twice & yes & once & no & no & no & no  & yes & yes & yes\\
duration (hr)                                        & $\sim4$ & $\sim2$ $>$2.5 & $\sim$6 & $\sim$4 & $\sim$7 & $>$3.3 & $\sim9$ & $\sim$12 & $\sim15$ & $\sim$14\\
$\tau_{\rm exp}$ (hr)                                & 0.7--1.9 & $\simeq$1 & 1.05--3.1 & 1.2$\pm$0.1 & 1.4$\pm$0.1 & 1.6$\pm$0.2 & $2.1$ & 2.7$\pm$0.1 & $\sim4$ & 6.0$\pm$0.3\\
$kT_{\rm max}$ (keV)                                 & 3.1$\pm$0.3 & $\simeq$3.0 & 2.35$\pm$0.01 & 2.6$\pm$0.2 & 2.6$\pm$0.2 & $\sim$2 & 1.5 & 2.4$\pm$0.1 & $2.1\pm 0.1$ & 1.8$\pm$0.1\\
$F_{\rm peak}$ (10$^{-8}$\,erg\,s$^{-1}$\,cm$^{-2}$)&1.5 & 7.0 & 2.0 & 1.9 & 1.5 & 3.5 & $>0.7$ & 2.3 & $\sim 4.5$ & 0.23\\
$L_{\rm peak}$ (10$^{38}$\,erg\,s$^{-1}$)$^c$  & $\simeq$1.8 & $\simeq$3.4 & $\simeq$1.3 & $\simeq$1.6 & $\simeq$1.5 & $\sim$0.8 & $\gtrsim0.09$ & $\simeq$1.4 & $0.4-0.7$ & $\simeq$0.4\\
$E_{\rm b}$ (10$^{42}$\,erg)                         & 0.3--1 & $\gtrsim$1.4 & $\simeq$0.65 & $\simeq$0.8 & $\gtrsim$0.5 & $\gtrsim$0.6 & $\gtrsim0.06$ & $\simeq$1.0 & $0.4-0.9$ & $\simeq$0.8 \\
$L_{\rm pers}$ (L$_{\rm Edd}$) & $\simeq$0.1     & $\simeq$0.8 & $\simeq$0.1 & $\simeq$0.2 & $\simeq$0.25 & $\sim$0.2 & $\simeq0.01$ & $\simeq$0.1 & $\simeq$0.1 & $\simeq$0.13 \\
H/He or He donor                                     & H/He & He & H/He & H/He & H/He & H/He & He?$^d$ & H/He & H/He & H/He\\
Orbital period                                       & ? & 11 min & 3.8 hr & ? & 4.7 hr & ? & 51 min & ? & 12.0 hr & 3.9 hr\\
\hline
\end{tabular}

\noindent
$^a$\,A question mark denotes an unknown value;

$^b$\,Numbers are based on Kuulkers et al. in preparation.

$^c$\,Unabsorbed bolometric peak (black-body) luminosity;

$^d$\,See \cite{2004Nelemans,Zand2005}

References: GX~17+2 \citet{Zand2004}, 4U~1820-303 \citet{Strohmayer2002},
4U~1636-536 \citet{Wijnands2001,Strohmayer2002a,2004Kuulkers}, Ser~X-1
\citet{Cornelisse2002}, 4U~1735-444 \citet{Cornelisse2000}, GX~3+1
\citet{Kuulkers2002}, 4U~0614+091 \citet{2005ATel..483....1K},
KS~1731-260 \citet{Kuulkers2002ks1731}, 4U~1608-522 \citet{2005ATel..482....1R,Keek2008},
4U~1254-690 \citet{Zand2003}.

\end{sidewaystable}
Superbursts are thought to originate in a 100 meter thick carbon-rich
layer containing the ashes of hydrogen and helium burning. Most superbursting
sources accrete hydrogen-rich material at a rate of $\dot{M}\gtrsim0.1\dot{M}_{\mathrm{Edd}}$
(e.g., \citealt{Kuulkers2003a}). The base of the carbon-rich layer
is close to the crust, which means that crustal heating is important
for reaching the ignition temperature for carbon burning. Therefore,
like intermediate bursts, superbursts are probes of the neutron star
crust.

For a superburst to occur, two main ingredients are required: sufficient
carbon to fuel the burst and a high enough temperature to ignite the
flash. Carbon is thought to be created mostly by stable helium burning, but it is
destroyed by the rp-process in the presence of hydrogen during bursts.
Since all known superbursters also exhibit normal bursts, this presents
a problem: how can sufficient carbon survive the many (at least hundreds)
normal X-ray bursts before a superburst? \citet{Keek2008} suggest
that sedimentation may separate carbon from the lighter elements hydrogen
and helium, but this should be investigated further. While normal
X-ray bursts pose the problem of removing carbon, this same process
aids in reaching the required ignition temperature. The heavy elements
that are created by proton captures on carbon, decrease the thermal
conductivity of the envelope. By reducing the efficiency of cooling,
this allows for the envelope to heat up faster, leading to shorter
superburst recurrence times of approximately one year (\citealt{Cumming2001}).
This leads to the realization that for most superbursters both stable
and unstable burning are required (\citealt{Zand2003,2003Schatz}).

The problems in understanding how these two ingredients are obtained
are well illustrated by the first observation of a superburst from
the `classical'%
\footnote{Previously, a superburst was observed from the long-duration transient
KS~1731-260 (\citealt{Kuulkers2002ks1731}). This source, however,
accreted at a high rate for over ten years before the superburst was
observed.%
} transient 4U~1608-522 (\citealt{Keek2008,2005ATel..482....1R}).
The neutron star started accreting at a high level only 55 days before
the occurrence of the superburst (Fig. \ref{fig:Superburst-from-4U}).
This is not enough time to accumulate the required carbon-rich layer,
so the layer must have been formed during multiple outbursts in the
previous 26 to 72 years. In the mean time, many normal type-I bursts
occurred. Furthermore, during accretion the crust heats up, but according
to the most recent models 55 days is not enough time to reach the
ignition temperature for a carbon flash. Likely, there is an important
extra source of heat in the crust, that is not taken into account
yet.

\section{On the possibility of detecting superbursts with INTEGRAL-ISGRI and
Swift-BAT}

While intermediate bursts have been detected with INTEGRAL-ISGRI and
Swift-BAT (e.g., \citealt{Chenevez2008} in these proceedings), no superbursts
have been reported from observations with these instruments. In this section
we investigate the possibility of detecting superbursts with IBIS/ISGRI and BAT.

IBIS has a field of view of the same order as the BeppoSAX WFCs. The
BeppoSAX instruments detected 4 superbursts in a typical exposure time
of 6 to 12 Msec per superburster (excluding the 4 superbursts from the
luminous source GX~17+2, \citealt{Zand2004}). The IBIS exposure times are of the
same order of magnitude. From this perspective one would expect a few
superburst detections in the ISGRI data. However, the $>15$ keV
bandpass of ISGRI is not ideal for the detection of few-keV black body
spectra. For a black body temperature of 3.0/2.5/2.0/1.5 keV, the
percentage of the photons above 15 keV is 10/5/1.5/0.3\% of the
total. For a bolometric flux of
$2\times10^{-8}$~erg~s$^{-1}$cm$^{-2}$, equivalent to roughly 1 Crab
between 2 and 10 keV, and a temperature of k$T$=2.5 keV, we expect an
IBIS photon countrate of 15 c/s on axis. This number is based on a
study of X-ray bursts in ISGRI data studied by \cite{Chelovekov2007}
and on time-resolved spectroscopy on such data by \cite{Molkov2005}
 and \cite{Chenevez2007}. If the same black body cools to
2.0 keV (following Stefan-Boltzmann), this decreases to 2.5 c/s.

We calculated ISGRI fluxes as a function of time for the superbursts
seen from 4U~1820-303, 4U~1636-536, Ser~X-1, 4U~1735-44, GX~3+1, 4U~1608-522
 and 4U~0614+091, using the numbers in Table \ref{tab:Overview-of-all}, and compared
that to the 15-40 keV light curves as provided by the ISDC. It turns
out that only 1 of the 7 superbursts could have been detected by
ISGRI: that from 4U~1820-303. The significance with respect to the
average error in the first superburst `science window' (an exposure of 1800 to 3600\,s) would have been
40 and the flux 7 times higher than the average persistent flux. These
high values are both due to the high peak temperature (3 keV) and the
large flux (2 times higher than the next brightest superburst). The other
superbursts would have been fainter than the persistent flux, and have
significances less than 4 per `science window'.

We performed a similar study for the BAT on Swift, and obtain a
similar conclusion. BAT is about twice less sensitive to X-ray
bursts. Only the superburst from 4U~1820-303 could have been detected,
with a significance of about 20 in the first superburst orbit. Other
superbursts have significances of 2 or less.

INTEGRAL has collected a large amount of exposure time on most (candidate) 
superbursters through a dedicated program to monitor the galactic center region (\citealt{Kuulkers2007}).
Although Swift does not have a similar program, it still has collected a
similar exposure time, since the field of view of BAT is eight times larger
than that of IBIS.
No superbursts are apparent from 4U~1820-303 in 15.6 Msec of ISGRI 
(from February 28, 2003 up to November 11, 2006) nor 10.4 Msec of BAT
 data (from February 12, 2005 up to November 11, 2008). Note that this is
consistent with the longer recurrence time that is predicted for pure helium
accretors (\citealt{Strohmayer2002}).

\section{Conclusion}

We reviewed the regimes of thermonuclear
burning of accreted matter on neutron stars. Unstable burning of hydrogen
and helium in the envelope is observable as type I X-ray bursts, which
allow us to study the neutron star envelope. Since a few years we
observe rare bursts with longer decay times of up to a few hours.
The so-called intermediate bursts are due to thermonuclear burning
of a thick helium layer, while the superbursts occur in a thick carbon-rich
layer. The base of these thick layers lies close to the crust, such
that the ignition of long bursts depends on the heating from the crust.
Therefore, intermediate bursts and superbursts can be used as probes
of the neutron star crust. The observation of one superburst --- the
first from a `classical' transient --- gives us important new insight
in the thermal properties of the crust. Observing more long duration
bursts will, therefore, allow us to improve and better constrain models
of crustal heating. The wide-field instruments
INTEGRAL-ISGRI and Swift-BAT have collected a large
amount of exposure time on many of the bursting sources in our Galaxy.
While these instruments have observed intermediate bursts, no superbursts
were detected yet. We find that due to the energy range of the instruments, only very bright superbursts may be detected.\\

\noindent \emph{Acknowledgements}. LK acknowledges support from The
Netherlands Organization for Scientific Research (NWO). 
We thank Erik Kuulkers for providing us with details of the superburst from 4U~0614+91 and for helpful comments.

\bibliographystyle{aa}
\bibliography{thesis_intro}

\begin{thebibliography}{51}
\expandafter\ifx\csname natexlab\endcsname\relax\def\natexlab#1{#1}\fi

\bibitem[{{Bildsten}(1998)}]{Bildsten1998}
{Bildsten}, L. 1998, in NATO ASIC Proc. 515: The Many Faces of Neutron Stars.,
  ed. R.~{Buccheri}, J.~{van Paradijs}, \& A.~{Alpar}, 419

\bibitem[{{Chelovekov} {et~al.}(2007){Chelovekov}, {Grebenev}, \&
  {Sunyaev}}]{Chelovekov2007}
{Chelovekov}, I.~V., {Grebenev}, S.~A., \& {Sunyaev}, R.~A. 2007, ArXiv
  e-prints, 709

\bibitem[{{Chenevez} {et~al.}(2008){Chenevez}, {Falanga}, {Kuulkers}, {Brandt},
  {lund}, \& {Cumming}}]{Chenevez2008}
{Chenevez}, J., {Falanga}, M., {Kuulkers}, E., {et~al.} 2008, ArXiv e-prints

\bibitem[{{Chenevez} {et~al.}(2007){Chenevez}, {Falanga}, {Kuulkers}, {Walter},
  {Bildsten}, {Brandt}, {Lund}, {Oosterbroek}, \& {Zurita
  Heras}}]{Chenevez2007}
{Chenevez}, J., {Falanga}, M., {Kuulkers}, E., {et~al.} 2007, \aap, 469, L27

\bibitem[{{Cooper} \& {Narayan}(2007)}]{Cooper2007a}
{Cooper}, R.~L. \& {Narayan}, R. 2007, \apj, 661, 468

\bibitem[{{Cornelisse} {et~al.}(2000){Cornelisse}, {Heise}, {Kuulkers},
  {Verbunt}, \& {in~'t~Zand}}]{Cornelisse2000}
{Cornelisse}, R., {Heise}, J., {Kuulkers}, E., {Verbunt}, F., \& {in~'t~Zand},
  J.~J.~M. 2000, \aap, 357, L21

\bibitem[{{Cornelisse} {et~al.}(2003){Cornelisse}, {in~'t~Zand}, {Verbunt},
  {Kuulkers}, {Heise}, {den Hartog}, {Cocchi}, {Natalucci}, {Bazzano}, \&
  {Ubertini}}]{Cornelisse2003}
{Cornelisse}, R., {in~'t~Zand}, J.~J.~M., {Verbunt}, F., {et~al.} 2003, \aap,
  405, 1033

\bibitem[{{Cornelisse} {et~al.}(2002){Cornelisse}, {Kuulkers}, {in't Zand},
  {Verbunt}, \& {Heise}}]{Cornelisse2002}
{Cornelisse}, R., {Kuulkers}, E., {in't Zand}, J.~J.~M., {Verbunt}, F., \&
  {Heise}, J. 2002, \aap, 382, 174

\bibitem[{{Cumming} \& {Bildsten}(2001)}]{Cumming2001}
{Cumming}, A. \& {Bildsten}, L. 2001, \apjl, 559, L127

\bibitem[{{Cumming} {et~al.}(2006){Cumming}, {Macbeth}, {in~'t~Zand}, \&
  {Page}}]{Cumming2006}
{Cumming}, A., {Macbeth}, J., {in~'t~Zand}, J.~J.~M., \& {Page}, D. 2006, \apj,
  646, 429

\bibitem[{{Eichler} \& {Cheng}(1989)}]{Eichler1989}
{Eichler}, D. \& {Cheng}, A.~F. 1989, \apj, 336, 360

\bibitem[{{Fisker} {et~al.}(2008){Fisker}, {Schatz}, \&
  {Thielemann}}]{Fisker2008}
{Fisker}, J.~L., {Schatz}, H., \& {Thielemann}, F.-K. 2008, \apjs, 174, 261

\bibitem[{{Fujimoto}(1993)}]{Fujimoto1993}
{Fujimoto}, M.~Y. 1993, \apj, 419, 768

\bibitem[{{Fujimoto} {et~al.}(1981){Fujimoto}, {Hanawa}, \&
  {Miyaji}}]{Fujimoto1981}
{Fujimoto}, M.~Y., {Hanawa}, T., \& {Miyaji}, S. 1981, \apj, 247, 267

\bibitem[{{Galloway} {et~al.}(2004){Galloway}, {Cumming}, {Kuulkers},
  {Bildsten}, {Chakrabarty}, \& {Rothschild}}]{1826:galloway04apj}
{Galloway}, D.~K., {Cumming}, A., {Kuulkers}, E., {et~al.} 2004, \apj, 601, 466

\bibitem[{{Galloway} {et~al.}(2006){Galloway}, {Muno}, {Hartman}, {Savov},
  {Psaltis}, \& {Chakrabarty}}]{Galloway2007}
{Galloway}, D.~K., {Muno}, M.~P., {Hartman}, J.~M., {et~al.} 2006, ArXiv
  Astrophysics e-prints

\bibitem[{{Gotthelf} \& {Kulkarni}(1997)}]{Gotthelf1997}
{Gotthelf}, E.~V. \& {Kulkarni}, S.~R. 1997, \apjl, 490, L161+

\bibitem[{{Grindlay} {et~al.}(1976){Grindlay}, {Gursky}, {Schnopper},
  {Parsignault}, {Heise}, {Brinkman}, \& {Schrijver}}]{Grindlay1976}
{Grindlay}, J., {Gursky}, H., {Schnopper}, H., {et~al.} 1976, \apjl, 205, L127

\bibitem[{{Gupta} {et~al.}(2007){Gupta}, {Brown}, {Schatz}, {M{\"o}ller}, \&
  {Kratz}}]{Gupta2007}
{Gupta}, S., {Brown}, E.~F., {Schatz}, H., {M{\"o}ller}, P., \& {Kratz}, K.-L.
  2007, \apj, 662, 1188

\bibitem[{{Haensel} \& {Zdunik}(2003)}]{Haensel2003}
{Haensel}, P. \& {Zdunik}, J.~L. 2003, \aap, 404, L33

\bibitem[{{in~'t~Zand}(2007)}]{Zand2007seattle}
{in~'t~Zand}, J.~J.~M. 2007, {Observations of rare and peculiar X-ray bursts},
  presented at "The Neutron Star Crust and Surface: Observations and Models" in
  Seattle.

\bibitem[{{in~'t~Zand} {et~al.}(2004{\natexlab{a}}){in~'t~Zand}, {Cornelisse},
  \& {Cumming}}]{Zand2004}
{in~'t~Zand}, J.~J.~M., {Cornelisse}, R., \& {Cumming}, A. 2004{\natexlab{a}},
  \aap, 426, 257

\bibitem[{{in~'t~Zand} {et~al.}(2005){in~'t~Zand}, {Cumming}, {van der Sluys},
  {Verbunt}, \& {Pols}}]{Zand2005}
{in~'t~Zand}, J.~J.~M., {Cumming}, A., {van der Sluys}, M.~V., {Verbunt}, F.,
  \& {Pols}, O.~R. 2005, \aap, 441, 675

\bibitem[{{in~'t~Zand} {et~al.}(2007){in~'t~Zand}, {Jonker}, \&
  {Markwardt}}]{intZand2007}
{in~'t~Zand}, J.~J.~M., {Jonker}, P.~G., \& {Markwardt}, C.~B. 2007, \aap, 465,
  953

\bibitem[{{in~'t~Zand} {et~al.}(2003){in~'t~Zand}, {Kuulkers}, {Verbunt},
  {Heise}, \& {Cornelisse}}]{Zand2003}
{in~'t~Zand}, J.~J.~M., {Kuulkers}, E., {Verbunt}, F., {Heise}, J., \&
  {Cornelisse}, R. 2003, \aap, 411, L487

\bibitem[{{in~'t~Zand} {et~al.}(2004{\natexlab{b}}){in~'t~Zand}, {Verbunt},
  {Heise}, {Bazzano}, {Cocchi}, {Cornelisse}, {Kuulkers}, {Natalucci}, \&
  {Ubertini}}]{2004ZandWFC}
{in~'t~Zand}, J.~J.~M., {Verbunt}, F., {Heise}, J., {et~al.}
  2004{\natexlab{b}}, Nucl. Phys. Proc. Suppl., 132, 486

\bibitem[{{Keek} {et~al.}(2008{\natexlab{a}}){Keek}, {in~'t~Zand}, {Kuulkers},
  {Cumming}, {Brown}, \& {Suzuki}}]{Keek2008}
{Keek}, L., {in~'t~Zand}, J.~J.~M., {Kuulkers}, E., {et~al.}
  2008{\natexlab{a}}, \aap, 479, 177

\bibitem[{{Keek} {et~al.}(2008{\natexlab{b}}){Keek}, {Langer}, \&
  {in~'t~Zand}}]{Keek2008a}
{Keek}, L., {Langer}, N., \& {in~'t~Zand}, J.~J.~M. 2008{\natexlab{b}}, in
  preparation

\bibitem[{{Kuulkers}(2002)}]{Kuulkers2002}
{Kuulkers}, E. 2002, \aap, 383, L5

\bibitem[{Kuulkers(2004)}]{Kuulkers2003a}
Kuulkers, E. 2004, Nucl. Phys. Proc. Suppl., 132, 466

\bibitem[{{Kuulkers}(2005)}]{2005ATel..483....1K}
{Kuulkers}, E. 2005, The Astronomer's Telegram, 483, 1

\bibitem[{{Kuulkers} {et~al.}(2002{\natexlab{a}}){Kuulkers}, {Homan}, {van der
  Klis}, {Lewin}, \& {M{\'e}ndez}}]{2002Kuulkers}
{Kuulkers}, E., {Homan}, J., {van der Klis}, M., {Lewin}, W.~H.~G., \&
  {M{\'e}ndez}, M. 2002{\natexlab{a}}, \aap, 382, 947

\bibitem[{{Kuulkers} {et~al.}(2004){Kuulkers}, {in~'t~Zand}, {Homan}, {van
  Straaten}, {Altamirano}, \& {van der Klis}}]{2004Kuulkers}
{Kuulkers}, E., {in~'t~Zand}, J., {Homan}, J., {et~al.} 2004, in AIP Conf.
  Proc. 714: X-ray Timing 2003: Rossi and Beyond, 257--260

\bibitem[{{Kuulkers} {et~al.}(2002{\natexlab{b}}){Kuulkers}, {in~'t~Zand}, {van
  Kerkwijk}, {Cornelisse}, {Smith}, {Heise}, {Bazzano}, {Cocchi}, {Natalucci},
  \& {Ubertini}}]{Kuulkers2002ks1731}
{Kuulkers}, E., {in~'t~Zand}, J.~J.~M., {van Kerkwijk}, M.~H., {et~al.}
  2002{\natexlab{b}}, \aap, 382, 503

\bibitem[{{Kuulkers} {et~al.}(2007){Kuulkers}, {Shaw}, {Paizis}, {Chenevez},
  {Brandt}, {Courvoisier}, {Domingo}, {Ebisawa}, {Kretschmar}, {Markwardt},
  {Mowlavi}, {Oosterbroek}, {Orr}, {R{\'{\i}}squez}, {Sanchez-Fernandez}, \&
  {Wijnands}}]{Kuulkers2007}
{Kuulkers}, E., {Shaw}, S.~E., {Paizis}, A., {et~al.} 2007, \aap, 466, 595

\bibitem[{{Lamb} \& {Lamb}(1978)}]{Lamb1978}
{Lamb}, D.~Q. \& {Lamb}, F.~K. 1978, \apj, 220, 291

\bibitem[{{Lewin} {et~al.}(1993){Lewin}, {van Paradijs}, \& {Taam}}]{Lewin1993}
{Lewin}, W.~H.~G., {van Paradijs}, J., \& {Taam}, R.~E. 1993, Space Science
  Reviews, 62, 223

\bibitem[{{Maraschi} \& {Cavaliere}(1977)}]{Maraschi1977}
{Maraschi}, L. \& {Cavaliere}, A. 1977, in Highlights in Astronomy, ed. E.~A.
  {M\"uller}, Vol.~4 (Reidel, Dordrecht), 127

\bibitem[{{Molkov} {et~al.}(2005){Molkov}, {Revnivtsev}, {Lutovinov}, \&
  {Sunyaev}}]{Molkov2005}
{Molkov}, S., {Revnivtsev}, M., {Lutovinov}, A., \& {Sunyaev}, R. 2005, \aap,
  434, 1069

\bibitem[{{Nelemans} {et~al.}(2004){Nelemans}, {Jonker}, {Marsh}, \& {van der
  Klis}}]{2004Nelemans}
{Nelemans}, G., {Jonker}, P.~G., {Marsh}, T.~R., \& {van der Klis}, M. 2004,
  \mnras, 348, L7

\bibitem[{{Peng} {et~al.}(2007){Peng}, {Brown}, \& {Truran}}]{Peng2007}
{Peng}, F., {Brown}, E.~F., \& {Truran}, J.~W. 2007, \apj, 654, 1022

\bibitem[{{Piro} \& {Bildsten}(2007)}]{Piro2007}
{Piro}, A.~L. \& {Bildsten}, L. 2007, \apj, 663, 1252

\bibitem[{{Remillard} {et~al.}(2005){Remillard}, {Morgan}, \& {The ASM Team at
  MIT}}]{2005ATel..482....1R}
{Remillard}, R., {Morgan}, E., \& {The ASM Team at MIT}, N. 2005, The
  Astronomer's Telegram, 482, 1

\bibitem[{{Schatz} {et~al.}(2003){Schatz}, {Bildsten}, {Cumming}, \&
  {Ouelette}}]{2003Schatz}
{Schatz}, H., {Bildsten}, L., {Cumming}, A., \& {Ouelette}, M. 2003, Nuclear
  Physics A, 718, 247

\bibitem[{{Strohmayer} \& {Bildsten}(2006)}]{Strohmayer2006}
{Strohmayer}, T. \& {Bildsten}, L. 2006, {New views of thermonuclear bursts}
  (Compact stellar X-ray sources), 113--156

\bibitem[{{Strohmayer} \& {Brown}(2002)}]{Strohmayer2002}
{Strohmayer}, T.~E. \& {Brown}, E.~F. 2002, \apj, 566, 1045

\bibitem[{{Strohmayer} \& {Markwardt}(2002)}]{Strohmayer2002a}
{Strohmayer}, T.~E. \& {Markwardt}, C.~B. 2002, \apj, 577, 337

\bibitem[{{van Paradijs} {et~al.}(1988){van Paradijs}, {Penninx}, \&
  {Lewin}}]{Paradijs1988}
{van Paradijs}, J., {Penninx}, W., \& {Lewin}, W.~H.~G. 1988, \mnras, 233, 437

\bibitem[{{Wijnands}(2001)}]{Wijnands2001}
{Wijnands}, R. 2001, \apjl, 554, L59

\bibitem[{{Woosley} \& {Taam}(1976)}]{Woosley1976}
{Woosley}, S.~E. \& {Taam}, R.~E. 1976, \nat, 263, 101

\bibitem[{{Yoon} {et~al.}(2004){Yoon}, {Langer}, \& {Scheithauer}}]{Yoon2004}
{Yoon}, S.-C., {Langer}, N., \& {Scheithauer}, S. 2004, \aap, 425, 217

\end{thebibliography}

\end{document}